\documentclass[10pt,letterpaper]{article}

\usepackage{opex3}
\usepackage{graphicx}
\usepackage{color}
\usepackage{cite}

\begin{document}

\title{Interaction of nematicons\\in a bias-free liquid crystal cell}

\author{Yana Izdebskaya$^1$, Vladlen Shvedov$^{1,2}$, Anton S. Desyatnikov$^1$,\\ Wieslaw Krolikowski$^2$, Gaetano Assanto$^3$, and Yuri S. Kivshar$^1$}

\address{
$^1$Nonlinear Physics Centre and $^2$Laser Physics Centre, Research School of Physics and Engineering, The Australian National University, Canberra ACT 0200, Australia\\
$^3$NooEL -- Nonlinear Optics and OptoElectronics Lab, Department of Electronic Engineering, CNISM--University of Rome ``Roma Tre'', Via della Vasca Navale 84, 00146 Rome, Italy}

\begin{abstract}
We study experimentally the propagation dynamics and interaction of spatial optical solitons in a bias-free cell filled with nematic liquid crystals. We reveal and measure long-range effects due to the cell boundaries for a single nematicon as well as for the interacting in-phase and out-of-phase nematicons. We discuss the effect of initial beam focusing and relative input angle on the interaction of in-phase nematicons.
\end{abstract}


\section{Introduction}

Solitons have been observed in diverse fields of nonlinear physics, and they share several common fundamental properties relying on a combination of nonlinear response and natural tendency of wave packets to spread as they propagate. Spatial optical solitons~\cite{Stegeman1999, Kivshar} originate from a balance between linear diffraction and self-focusing in a nonlinear optical medium, and they are self-confined (non-spreading) beams, i.e. their size is invariant during propagation. Spatial optical solitons have been investigated extensively in a variety of nonlinear materials, both in one- and two-dimensional geometries; they have significant potentials towards signal processing, switching and readdressing in future generations of all-optical circuits. In this context, a giant nonlinearity arising from molecular reorientation in nematic liquid crystals (NLCs) attracted significant attention~\cite{Tabirian, Khoo, Simoni}. Both experimental~\cite{Peccianti2000, Hutsebaut} and theoretical~\cite{McLaughlin1996} results were demonstrated for spatial solitons in NLCs, also known as {\em nematicons}~\cite{Assanto_OPN}.

Nematic liquid crystals consist of elongated molecules aligned along a given direction (molecular director) owing to both anchoring at the boundaries and intermolecular forces~\cite{Tabirian, Khoo, Simoni}. The resultant medium is positively birefringent uniaxial with ordinary and extraordinary refractive indices, $n_{\|}$ and $n_{\perp}$, defined for polarizations parallel and orthogonal to the director. In general, the properties of NLCs are intermediate between solids and liquids, thus liquid crystals can be found in different phases, featuring different properties and degree of order. Their giant reorientational nonlinearity allows to generate nematicons with relatively low optical powers, at the milliWatt level or below, for the study of the fundamental aspects and applications of light interactions with self-assembling soft matter.

\begin{figure}
\centering\includegraphics[width=10cm]{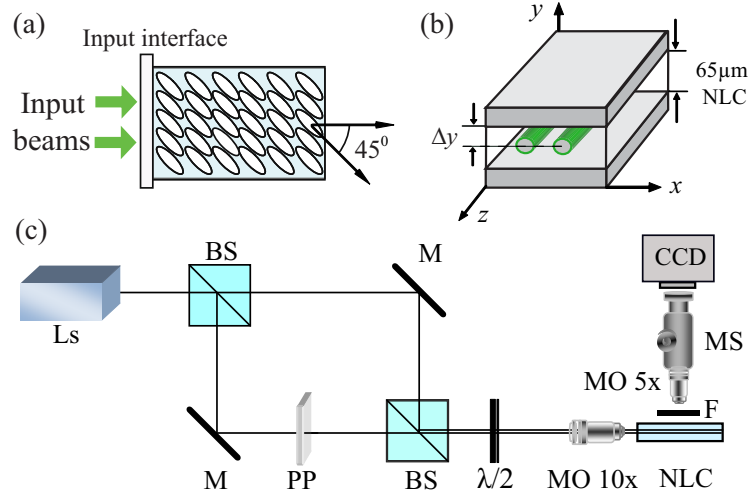}
\caption{(a) Top view of the NLC planar cell. The ellipses indicate the orientation of the molecules in the plane $(y,z)$. (b) Perspective view of the NLC cell with two nematicons.  (c) Experimental setup: Ls -- cw-laser ($\lambda=532nm$), BS -- beam splitters, PP -- parallel face plate for adjusting the phase difference, M -- mirrors, $\lambda/2$ -- half wavelength plate, MO -- microscope objectives, NLC -- nematic liquid crystal cell, F -- filter, MS -- microscope, CCD -- camera.}
\label{fig1}
\end{figure}

While earlier studies of solitons in nematic liquid crystals considered the propagation of a single nematicon~\cite{Peccianti2000, Peccianti2001, Conti2003, Assanto2003, Conti2004, Peccianti2004, Alberucci2005}, the interactions of two nematicons of the same~\cite{Peccianti2002,Fratalocchi2004,Fratalocchi2007,Fratalocchi2007PRE,Smyth2008} or different wavelengths~\cite{Alberucci2006, Assanto2008} were addressed more recently. The power- and time-dependent attraction and fusion of nematicons in bulk undoped NLC in planar voltage-biased liquid-crystal cell were observed experimentally by Peccianti {\it et al.}~\cite{Peccianti2002, Peccianti2002_APL}, revealing that nonlocality plays a key role for their long-range phase- and wavelength-independent interactions. Hu {\it et al.}~\cite{Hu2006, Hu2008} underlined the differences between the short- and long-range interactions of two nematicons: at variance with the short-range phase-sensitive interactions, the long-range attraction dominates in strongly nonlocal media.

Several experiments with nematicons were carried out in NLC cells with an applied (low frequency) electric field (voltage), which can control the orientation of the liquid crystal molecules and, therefore, their nonlinear as well as nonlocal responses~\cite{interplay}, thus affecting the interaction between nematicons. Nematicons can also propagate in unbiased NLC cells~\cite{Peccianti2004, Alberucci2005, Alberucci2006, Fratalocchi2007, Fratalocchi2007PRE, Piccardi2008}, as well as in bias-free chiral and twisted NLC cells~\cite{Laudyn2009, PLP}.

In this paper we study experimentally the propagation of a single nematicon and the interaction between two nematicons in different layers of a planar un-biased NLC cell in order to assess the role of the boundaries, previously investigated numerically by Alberucci {\it et al.}~\cite{AlberucciJOSAB}. We observe both phase-sensitive interaction of nematicons, similar to the case of local nonlinear media, and suppression of out-of-phase soliton repulsion, a signature of long-range interactions. The co-existence of two effects pinpoints an intermediate range of nonlocality in bias-free NLCs.

\section{Experimental results}

We adopt the cell geometry sketched in Fig.~\ref{fig1}(a), with two parallel polycarbonate plates separated by a gap, the latter defined  by spacers of  thickness $65\mu m$. The cell contains the 6CHBT liquid crystals~\cite{Baran, Dabrowski} which has negligible absorption and high nonlinearity with refractive indices $n_e$=1.6718 and $n_o$=1.5225 at room temperature. The polycarbonate plates have been rubbed in the plane $(x,z)$ at an angle $\pi/4$ with respect to $z$. Such boundary conditions entail molecular orientation in bulk analogous to that provided by  pre-alignment via an external bias. However, in contrast to a standard biased NLC cell where the director can rotate orthogonally to the plates in the plane $(y,z)$, the optic axis here remains  parallel to the rubbed plates and the light for soliton excitation is polarized along the $x$ direction, see Fig.~\ref{fig1}(b). Using a $10\times$ microscope-objective, a Gaussian beam of power 4.7~mW is focused onto the cell input facet. A standard half-wave plate controls the polarization and allows choosing between linear (ordinary wave) or nonlinear (extraordinary wave) regimes of propagation. The beam evolution is monitored by collecting the light scattered above the cell (from the plane $(x,z)$) with an optical microscope, a $5\times$ microscope-objective and a high resolution CCD camera. The input beam waist is $w_0=3.2\,\mu$m, i.e. its transverse size is well below the cell thickness. We keep the total beam power at low values in order to avoid thermooptical effects arising from absorption and heating.

\begin{figure}
\centering\includegraphics[width=13cm]{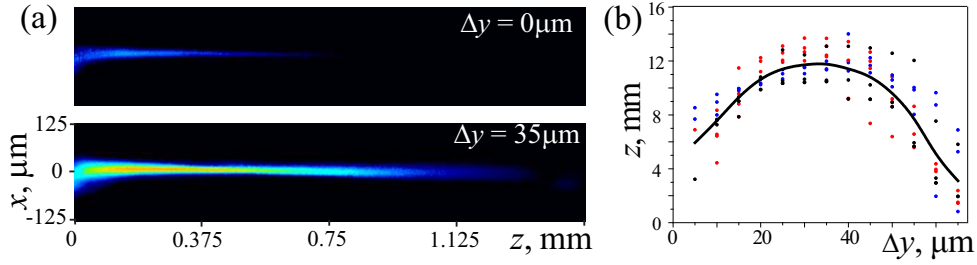}
\caption{(a, {\color{blue} Media 1}) Experimentally recorded images of the light scattered from a single nematicon at different positions $\Delta y$ in the cell relative to the upper plate; the input beam power is 4.7~mW. (b) Averaged distance (solid line) at which the intensity of the scattered radiation is at its half-maximum; data points are shown by colored dots in black for a single nematicon, in red for two in-phase and in blue for two out-of-phase nematicons.}
\label{fig2}
\end{figure}

We begin by studying the longitudinal dynamics of a single nematicon versus its position $\Delta y$ with respect to the cell upper boundary, see Fig.~\ref{fig1}(a, {\color{blue} Media 1}). In order to record the beam trajectories in the plane $(y,z)$ parallel to the cell plates and to control the vertical position of the soliton in the NLC cell, we use an $(x,y,z)$-axis flexure stage and gradually move the cell (with increments of $5\,\mu$m) along $y$ from top to bottom. As seen in Fig.~\ref{fig2}(a), self-focusing at the given power is sufficient for beam self-trapping at any position in the cell, confirming the generation of the bias-free nematicons.

However, the visible propagation length of the nematicon changes dramatically with $\Delta y$, attaining a maximum in the midplane of the cell. The reason for this is the different amount of light scattering by various NLC layers, with higher degree of orientational order (lower scattering) closer to the boundaries. As apparent in Fig.~\ref{fig2}(b), in the  region $15\,\mu$m$<y<50\,\mu$m in the cell the ``effective length'' of the nematicon is nearly constant.

\begin{figure}
\centering\includegraphics[width=13cm]{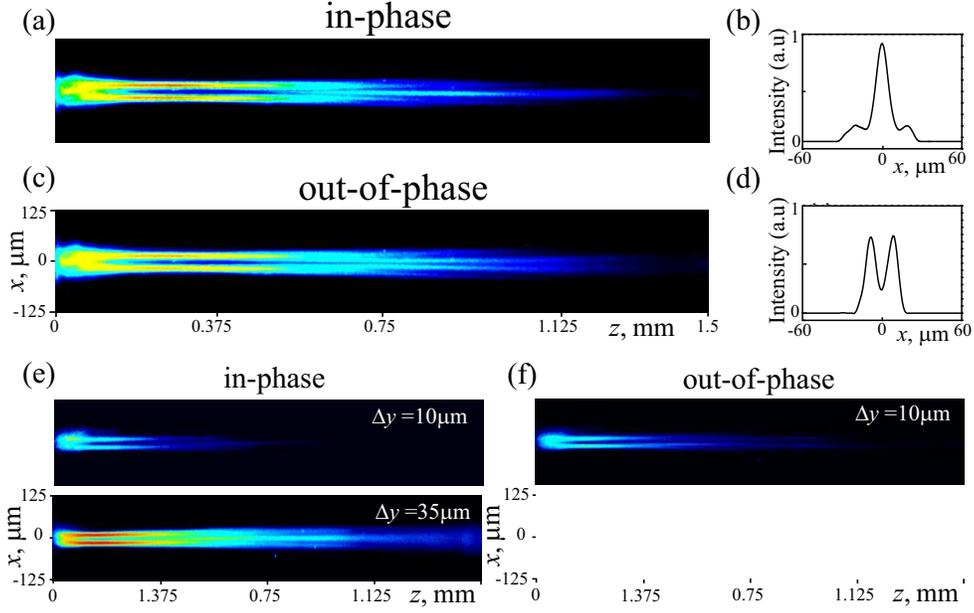}
\caption{Typical intensity distributions for the interacting in-phase (a, b) and out-of-phase (c, d) nematicons. The input power of each beam is 2.35~mW. The profiles (a-d) recorded for the beams in the middle of the cell, $\Delta y=32.5\,\mu$m; (b) at $z=0.85$~mm and (d) at $z=1.2$~mm. Similar to Fig.~\ref{fig2} we also compare the profiles for both the in-phase (e, {\color{blue} Media 2}) and out-of-phase (f, {\color{blue} Media 3}) nematicons versus their position in the cell $\Delta y$.}
\label{fig3}
\end{figure}

Next we consider the interaction of two nematicons in the same cell. In order to generate two identical input beams we use the standard Mach-Zehnder arrangement, see Fig.~\ref{fig1}(c). The relative phase between the two beams is adjusted by the rotation of a 1.8~mm-thick parallel-face plate (PP) and is measured via the interference pattern on a  beam profiler after the second beam-splitter. With the half-wave plate we control the polarization state of the two identical parallel Gaussian beams of power 2.35~mW each, focused into the cell by a $10\times$ objective. Figure~\ref{fig3} shows the experimental results for the interaction between  in-phase and out-of-phase nematicons. The waist of each beam is $w_0=3.2\,\mu$m and the distance between them is $12\,\mu$m.

As it was discussed earlier~\cite{Snyder1997, Peccianti2002, Rasmussen2005, Hu2006}, the long-range interaction between solitons is largely independent on their relative phase, similar to the interaction of mutually incoherent solitons~\cite{Peccianti2002, Peccianti2002_PRE, Shen2005, Ku2005}. Conversely, the phase-dependence in  short-range interaction between nematicons~\cite{Hu2006, Hu2008} is reminiscent of those in local nonlinear media~\cite{Stegeman1999, Kivshar}. On one hand we observe clear differences between in-phase and out-of-phase interactions, cf. Figs.~\ref{fig2}(a, b) and (c, d); in particular, in-phase nematicons strongly exchange power and merge in (a, b). On the other hand  out-of-phase nematicons as in Figs.~\ref{fig2}(c, d) remain almost parallel, i.e. the mutual repulsion, expected for interactions in local media, is suppressed by nonlocality. We conclude that in our experiments an intermediate regime is probed, where the nonlocality is high enough to balance out-of-phase repulsion without completely suppressing the phase-dependent behavior.

To study the role of cell boundaries on nematicon interactions we vary the position $\Delta y$ of the nematicon pair relative to  the cell upper plate. As visible in Fig.~\ref{fig3}(e, {\color{blue} Media 2}) for in-phase nematicons, and in Fig.~\ref{fig3}(f, {\color{blue} Media 3}) for out-of-phase nematicons, there is no qualitative difference in the interaction between nematicons positioned anywhere in the NLC thickness, except for small regions by the boundaries.

\begin{figure}
\centering\includegraphics[width=13cm]{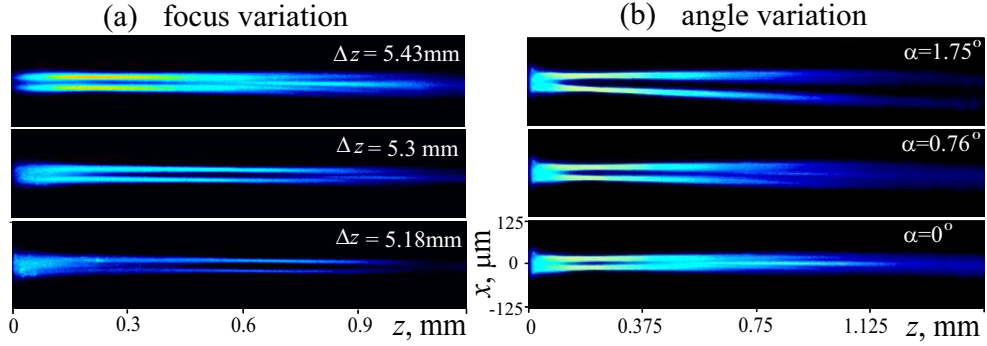}
\caption{Experimental results on the interaction between two nematicons. (a, {\color{blue} Media 4})  Effects due to varying the $z-$position of the focal plane of the input objective lens (5.5~mm) with respect to the input facet; here $\Delta z$ is the distance between the objective and the NLC cell. (b, {\color{blue} Media 5})  Effects due to variations in the launch angle $\alpha$ between two in-phase nematicons.}
\label{fig4}
\end{figure}

Finally, Figs.~\ref{fig4}(a, b) summarize our experimental results on the strength of the coherent interaction between  in-phase nematicons versus launch conditions, i.e. input focusing and relative angle between the beams. In both cases the nematicons are excited in the midplane. By shifting the focal plane of the input  microscope-objective within the cell [Fig.~\ref{fig4}(a, {\color{blue} Media 4})], we observe a shift of the interaction region along $z$ until it seemingly vanishes for $\Delta z< 5.18$~mm, i.e. the focal plane in $z>0.32$~mm. In other words, an additional focusing reduces the efficiency of the power transfer between nematicons. We also find a threshold in the angle $\alpha$ between the two input beams [Fig.~\ref{fig4}(b, {\color{blue} Media 5})]: below it the interaction results in nematicon coalescence, whereas above it the nematicons practically do not interact. The threshold is $\alpha\simeq 0.76^\circ$.

\section{Conclusions}

We have studied experimentally the generation and phase-sensitive interaction of two nematicons in an unbiased nematic liquid  crystal cell. The boundary effects arising from anchoring of the NLC can be estimated by the reduction of the scattering as the soliton gets closer to the rubbed surface. For cells $65\,\mu$m thick, a central region of $\sim35\,\mu$m in the cell provides an almost constant nonlinear response. Furthermore, we have observed short-range phase-sensitive nematicon interactions along with nonlocality-driven suppression of the repulsion between out-of-phase solitons.

All the results involving the role of finite boundaries and phase-sensitive soliton-soliton interactions encourage the investigation of more complex multi-soliton states. These include boundary-induced motion of nematicons~\cite{PRA} as well as intriguing self-induced transformations between higher-order nematicons~\cite{OE,JOA}.

This work was supported by the Australian Research Council. The authors thank E. Brasellet for stimulating discussions.

\end{document}